# A Trust Model Based on Service Classification in Mobile Services

Yang Liu, Zhikui Chen, Feng Xia, Xiaoning Lv and Fanyu Bu
School of Software
Dalian University of Technology
Dalian 116620, China
E-mail: lastyang1984@hotmail.com

*Abstract*—Internet of Things (IoT) and B3G/4G communication are promoting the pervasive mobile services with its advanced features. However, security problems are also baffled the development. This paper proposes a trust model to protect the user's security. The billing or trust operator works as an agent to provide a trust authentication for all the service providers. The services are classified by sensitive value calculation. With the value, the user's trustiness for corresponding service can be obtained. For decision, three trust regions are divided, which is referred to three ranks: high, medium and low. The trust region tells the customer, with his calculated trust value, which rank he has got and which authentication methods should be used for access. Authentication history and penalty are also involved with reasons.

## I. INTRODUCTION

With the rapid promotion, the Internet of Things (IoT) and B3G/4G communication system will offer us great convenience and huge opportunities of pervasive [1, 2] mobile service creation. But due to its wider application providing for billions of users, many security threats are also accompanied. Such as identity theft, malicious or illegal service request, the loss or disclosure of personal data, privacy and related intellectual property, user identity authentication is an attractive issue that must be thoroughly addressed. However, some existing revolutions are so complex that bringing bad influence on service quality and the consumers' convenience. Trust scheme seems to be a balanced way in this aspect.

In the human society, the social interactions are built around trust. The interaction histories or the evaluations from others are used to build the reputations of each other. In [3], the properties of trust are summarized as: subjectivity, non-transitivity, temporalness, contextualness and dynamicity, as well as non-monotonicity. There are numerous different kinds of trust that satisfy different properties that can be established differently. In terms of computer science, there are many definitions and models for trust. As pointed in [4], based on the trust values of transactions recently, a trust evaluation approach is proposed for e-commerce applications. The trust values are random samples. In this method, recent trust values are more important in the trust evaluation. The approach in [5] of the fuzzy logic is applied to trust evaluation, which divides sellers or service providers into multiple classes of reputation ranks. A model for supporting trust in virtual communities is proposed in [6]. It is based on direct experiences and reputation. For agent system, trust management is also actively proposed, such as [7, 8]. In Daidalos I and II, a Virtual Identity's concept [9] is designed to protect a user's privacy and secure communication data, which contemplates the multitude of identities and roles we take on each time we turn on our computer, mobile phone or PDA. The user has a contract with the trusted operator, who becomes a proxy for billing, which is a business in itself.

We accept these methods' effectiveness and make an integrated scheme to classify the user's trust rank and service sensitivity. We bring trust rank in three levels: high, medium and low. For each rank, the authentication way is varied. In high rank case, No extra key is needed. For medium rank, users have to offer their PIN for login. Low rank means users need to provide the biometric information, such as face image, fingerprint and iris scan, which may be not convenient for its complexity and hardware constraint. Then, the authentication history will also affect the access. In addition, in order to protect the force-crack by attackers, we induce a penalty coefficient, which the trying times are extremely limited.

The remainder of the paper is organized as follow. Section 2 shows details of services classification. Section 3 proposes a trust evaluation model. Simulations are made in Section 4. Section 5 concludes the paper.

## II. SERVICES CLASSIFICATION

Our system bases service classification on static information about the applications, such as the type of application, the cost of service and the host on which the application was executed. As a consequence of this split of multiple types of services are supported in the trust operator. The classification is basically related with the sensitivity or importance of the service provider. Here, we exploit the fuzzy mathematics to quantity the sensitivity.

### A. Generate comparison matrix

Firstly, we make pairwise comparison of the importance of n targets $X_1, X_2, \ldots, X_n$ presenting the services. According the combination principle, there are $1/2 \times n \times (n-1)$ times comparisons. The $X_i$ and $X_j$ are evaluated by the relative importance using relative

importance value table, see table 1, which can be labeled as $a_{ij}$. Hence, we have:

$$A = \begin{bmatrix} a_{11} \dots a_{1n} \\ \dots\dots\dots \dots\dots \\ a_{n1} \dots a_{nn} \end{bmatrix} = (a_{ij})_{n \times n}$$

where $a_{ij} = f(X_i, X_j)$ presenting the importance of $X_i$ compared with $X_j$.

TABLE I.     RELATIVE IMPORTANCE VALUE TABLE

| Comparison between service *i* and *j* | $f(X_i,X_j)$ | $f(X_j,X_i)$ |
|---|---|---|
| *i* equal important to *j* | 1 | 1 |
| *i* slightly more important than *j* | 3 | 1/3 |
| *i* obviously more important than *j* | 5 | 1/5 |
| *i* great more important than *j* | 7 | 1/7 |
| *i* extremely more important than *j* | 9 | 1/9 |
| *i* and *j* in the intervals of two judgement | 2,4,6,8 | 1/2,1/4,1/6,1/8 |

### B. Confirm the weight coefficients

For comparison matrix $A = (a_{ij})_{n \times n}$, the approximate eigenvalue and units eigenvector can be calculated by rooting, these targets of weighted vector $\overline{W}$:

*i.* Calculate $a_i = \sqrt[n]{\prod_{j=1}^{n} a_{ij}}$, $i,j = 1,\dots,n$, in row of A.

*ii.* Normalize $a_i$ as $W_i = a_i / \sum_{i=1}^{n} a_i$, $i, j = 1,\dots,n$.
$\overline{W} = (W_1, W_2, \dots, W_n)^T$ is approximate eigenvector of matrix A.

*iii.* Calculate $\lambda_{\max} = \frac{1}{n} \sum_{i=1}^{n} \frac{(A_i \times \overline{W})}{W_i}$ as approximate largest eigenvalue of matrix A.

### C. Self-consistency validation

Since $a_{ij} = f(X_i, X_j)$ is a kind of subjective evaluation, we have to make sure that our estimation of $a_{ij}$ does not contradict one another. For example, the value K is more important than L, and L is more important than M, then our evaluation function supports the conclusion that K is more important than M. If we value K is equal important to M, we may well doubt whether the evaluation function is reasonable. Self-consistency validation is a way to solve this problem.

Here, we calculate $CI = \frac{\lambda_{\max} - n}{n - 1}$, here $n$ is the order of matrix $A$. Then, consistency ratio $CR = \frac{CI}{RI}$, here $RI$ presents Average Random Consistency Index, see table 2. If CR＜0.1 we can accept the evaluation matrix $A$ as well as the weight vector $\overline{W}$. However, if CR≥0.1 there may exists conflict in the evaluation function. So we have to adjust matrix $A$ and do all the steps above again.

TABLE II.     STYLESAVERAGE RANDOM CONSISTENCY INDEX [10]

| Order | 1 | 2 | 3 | 4 | 5 | 6 | 7 | 8 |
|---|---|---|---|---|---|---|---|---|
| RI | 0 | 0 | 0.52 | 0.89 | 1.12 | 1.26 | 1.36 | 1.41 |
| Order | 9 | 10 | 11 | 12 | 13 | 14 | 15 | |
| RI | 1.46 | 1.49 | 1.52 | 1.54 | 1.56 | 1.58 | 1.59 | |

### D. Matrix calculation

The comparison matrix can be attached with some reasonable values. And as described in section 2(B), we have table 3 and table 4.

TABLE III.     MATRIX CALCULATION (1)

| $a_{ij}$ | A | B | C | D | E | F | G | H | I |
|---|---|---|---|---|---|---|---|---|---|
| A | 1 | 2 | 3 | 4 | 5 | 6 | 7 | 8 | 9 |
| B | 1/2 | 1 | 2 | 3 | 4 | 5 | 6 | 7 | 8 |
| C | 1/3 | 1/2 | 1 | 2 | 3 | 4 | 5 | 6 | 7 |
| D | 1/4 | 1/3 | 1/2 | 1 | 2 | 3 | 4 | 5 | 6 |
| E | 1/5 | 1/4 | 1/3 | 1/2 | 1 | 2 | 3 | 4 | 5 |
| F | 1/6 | 1/5 | 1/4 | 1/3 | 1/2 | 1 | 2 | 3 | 4 |
| G | 1/7 | 1/6 | 1/5 | 1/4 | 1/3 | 1/2 | 1 | 2 | 3 |
| H | 1/8 | 1/7 | 1/6 | 1/5 | 1/4 | 1/3 | 1/2 | 1 | 2 |
| I | 1/9 | 1/8 | 1/7 | 1/6 | 1/5 | 1/4 | 1/3 | 1/2 | 1 |

TABLE IV.     MATRIX CALCULATION (2)

| Value | $a_i$ | $W_i$ | $A_i\overline{W}$ |
|---|---|---|---|
| A | 4.14716627 | 0.30941616 | 2.91617 |
| B | 3.00799234 | 0.22442346 | 2.07087 |
| C | 2.11309937 | 0.15765635 | 1.44093 |
| D | 1.4592328 | 0.10887198 | 0.99041 |
| E | 1 | 0.07460905 | 0.6759 |
| F | 0.68529161 | 0.05112896 | 0.46075 |
| G | 0.47323851 | 0.03530788 | 0.31705 |
| H | 0.33244766 | 0.02480361 | 0.22451 |
| I | 0.18473035 | 0.01378256 | 0.1375 |

Then we get:

$$\lambda_{\max} = 9.2185899, CI = 0.0273237, CR = 0.0187149 < 0.1$$

So pass the check.

We can have the initial service classification estimation in table 5.

TABLE V. ESTIMATION TABLE

| Level | Sensitive value | Services |
|---|---|---|
| A | $S_1$=0.30941616 | Governmental\military |
| B | $S_2$=0.22442346 | Commercial |
| C | $S_3$=0.15765635 | Academic |
| D | $S_4$=0.10887198 | Banking\Stork |
| E | $S_5$=0.07460905 | e-Shopping |
| F | $S_6$=0.05112896 | VoIP |
| G | $S_7$=0.03530788 | Education |
| H | $S_8$=0.02480361 | Entertainment |
| I | $S_9$=0.01378256 | Public |

If a new service is emerging, of course, the table needs to be updated. The only task we should do is to make comparison between new service and original services and give a subjective value. For example, the new service is about online payment and initial estimate is between level D and E. Then, the matrix is expanded with subjective comparison value. Finally, the arrangement of levels and services are also renewed.

III. TRUST EVALUATION MODEL

A. Trust region

Three trust regions are divided representing three ranks. Each rank has its own entering way. In high rank case, No extra key is needed (already sign on the VID). For medium rank, users have to offer their PIN for login. Low rank means users need to provide the biometric information, such as face image, fingerprint and iris scan. To distinguish the rank of a specified field, an upper limitation $\Omega$ and a lower limitation $\omega$ is approximately defined for each certain field according to variable requirement, where $0 \leq \omega \leq 0.5 < \Omega \leq 1$. Therefore, three sets, $[0, \omega)$, $[\omega, \Omega)$ and $[\Omega, 1]$, are defined to calculate the thresholds of that field. The thresholds selection is related with the service classification.

Initially, the two thresholds can be set by the customer via choosing the sensitive values of services. The operator provides the recommended value. For example, the customer considers the upper threshold and lower threshold 0.7 and 0.3 respectively. Therefore, the three regions are [0, 0.3), [0.3, 0.7) and [0.7, 1]. The Trust Region can be verified by the Authentication History Value. The details will be explained later.

B. Authentication History

The customer's authentication history is also an important factor for reference. Based on past experience in authenticating consumer, the trust provider will make statistic for the situation of authentication. A nice history will make the customer's senior trust region much easier to access. In contrary, bad history will bring stricter ways for validate. Due to the three rank exploit different validation methods, the statistic models are distinguished.

The statistic results are labeled by $T_1$, $T_2$ and $T_3$ for high rank, medium rank and low rank respectively. Since high rank needs no extra key, this authentication can certainly success. Therefore, $T_1$ is the ratio of high-rank login to the totality. It says if $T_1$ is high, the customer who has an extremely regular life usually success in access with no PIN. Naturally, the user certainly is treated as trustiness. If T1 is not high, he may be not easy to be recognized, perhaps not in regular. $T_2$ is the successful PIN login radio. If $T_2$ is high, it is deduced his life may be less regular but highly PIN-control. But when $T_2$ is low, to some extent, the user often makes mistakes in PIN or lost PIN control, and may be less trusted. Since the biometric information is the private feature that can be highly trusted, if the failure happens, it says the user can not be trusted. While biometric check is complex, no one can guarantee the computers can recognize the features with only one chance or two and also the features have small possibility to loose. Thus, $T_3$ can be ignored here. In particular case, for $T_2$, if the history data of the customer with a good record in the past becomes bad currently, it can estimate that the identity is not safe this period, for example, is stolen.

With authentication history involving, the trust thresholds have some adjustment. We have:

$$\omega' = \omega + a. \quad (1)$$

$$\Omega' = \Omega + b. \quad (2)$$

Where $a$ and $b$ are function of $T_2$ and $T_1$, and also related with two region, $\Omega - \omega$ and $1 - \Omega$ respectively, which means $a$ and $b$ can not exceed the bounds. We simply choose function of incremental curves here. In addition, due to high rank is enough sensitive and constrained, it can not be changed obviously as that of medium can so that the curve should be much more steady. Therefore, $a$ and $b$ are selected with experience and test in (3) and (4), illustrated as Fig. 1.

$$a = \frac{e^{1-T_2} - 1}{e - 1} \times (\Omega - \omega). \quad (3)$$

$$b = \frac{e^{1-T_1} - 1}{e + 1} \times (1 - \Omega). \quad (4)$$

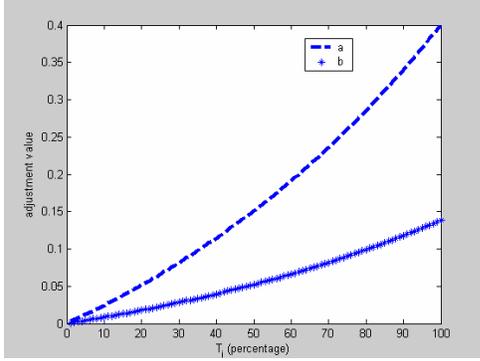

Figure 1. The forms of a and b.

## C. Trust evaluation

We map $S_i$ to [0,1] region with logarithm function as curve fitting:

$$S' = \frac{\lg S_i - \lg(\min S_i)}{\lg(\max S_i) - \lg(\min S_i)}. \quad (5)$$

And if $S_i$ is higher, the trust estimate should be more rigorous. Therefore, for trust value, we make an inverse of $S_i$:

$$Y^* = 1 - S'. \quad (6)$$

The system can judge that which region is matched by $Y^*$, according the thresholds. And then it will decide which authentication key should be provided by the customer. In order to counteract the improper setting of thresholds by the customers, a calibration factor $\frac{\Omega + \omega + 1}{2}$ is involved. For instance, the upper threshold is 0.8 and lower threshold is 0.5, which seems the set is a little high. $\frac{\Omega + \omega + 1}{2} = 1.15$ makes the trust value increase along and gently offsets the influence. Hence, (5) can be revised to:

$$Y = Y^* \times \frac{\Omega + \omega + 1}{2}. \quad (7)$$

For example, the service provider is an online market and the expense is not high, which selects the sensitive value $S = 0.10887198$ based on the sensitive value table 4. Upper threshold and lower threshold are 0.7 and 0.3 respectively. Here, $Y = 0.45718$. The trust region according to user's choice and authentication history is [0, 0.3245), [0.3245, 0.7663) and [0.7663, 1] ($T_1 = 0.4$ usually not every high and $T_2 = 0.9$ for simple, using (3) and (4)). Obviously, it is in the medium rank and PIN is necessary for authentication. When a user wants to enter into military services, according to (5), the trust value is zero in any case so the bio-information is required definitely.

## D. Penalty coefficient

We do not hope the theft try the PIN with no constraint. So the Penalty Coefficient is defined:

$$Y' = YP^n. \quad (8)$$

$n$ is the times of failure. If $Y$ belongs to medium region for example, the user can authenticate with PIN. But after several failures, $P^n$ becomes small enough, the $Y'$ may fall into low region and PIN is useless now.

The Penalty Coefficient can be set by calculation. In the first transaction, for medium rank, as $n$ times wrong in entering, the trust region falls into the low rank, which means the upper threshold decreases to lower threshold. Hence,

$$\omega = \Omega P^n. \quad (9)$$

While the customer chosen the threshold and trying times, the Penalty Coefficient can be established. For instance, $n = 5$, we have $P = 0.844$ according to (9). Then, for 1 failure in the example above, we get $Y' = 0.3895$ from (8), which still locates in medium rank. For second failure, $Y' = 0.3257$, also in medium rank. But for third failure, $Y' = 0.2749$ and trust value falls to low rank.

## IV. SIMULATION

### A. Experiment 1

We test upper threshold's effect to the Trust Value.

*i.* Without the authentication history involved, the lower threshold $\omega$ is fixed at 0.3. We select $S_1 = 0.1577$, $S_2 = 0.0353$ and $S_3 = 0.0248$ as samples. All the three curves climb slightly when the upper thresholds increase, seen from Fig.2(a). And for each upper threshold, the trust value follows $Y_3 > Y_2 > Y_1$. The red circle dots for $S_3$ and red diamond dots for $S_2$ present the trust values bigger than the upper thresholds, which means at this time the customer is in the high rank. It implies the higher the sensitivity of service is, the harder the high rank is entered.

*ii.* With consideration of authentication history, the other parameters unchanged as above. $T_1 = 0.4$ and $T_2 = 0.9$ are assumed here. The rank situation shows as Fig.2(b). Obviously, with the authentication history, comparing with Experiment 1(i), the high trust region for $S_3$ is reduced and for $S_2$ is even deleted. Additionally, the lower threshold is increasing with upper threshold, which makes the medium region also smaller. It demonstrates that the authentication history makes the scheme much stricter.

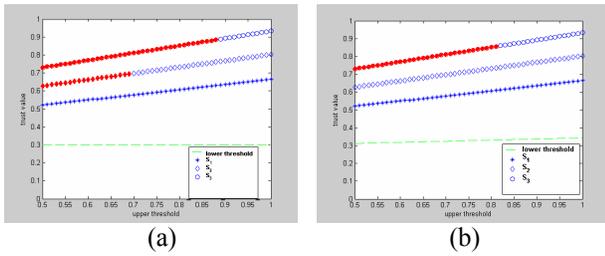

(a)  (b)

Figure 2. Upper threshold's effect: (a) Without authentication history (b) With authentication history.

As (6), lower threshold also affect the trust value and appears similar with upper threshold.

*B. Experiment 2*

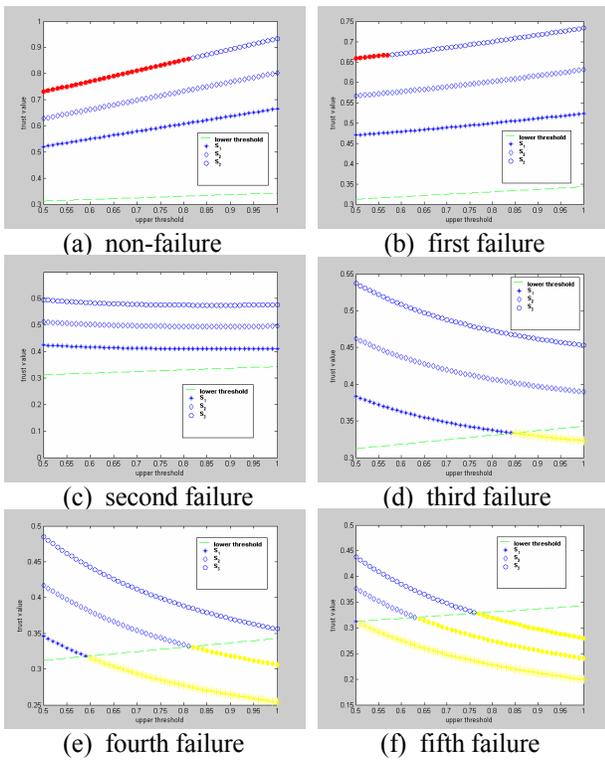

(a) non-failure  (b) first failure
(c) second failure  (d) third failure
(e) fourth failure  (f) fifth failure

Figure 3. Situation of penalty coefficient involved

In this experiment, we base the experiment 1(ii) and discuss the penalty coefficient. As the example in section 3(D), we set that the times of trials PIN is up to 5. The penalty coefficient $P$ is also related with $\Omega$ varying. The trust value and rank are changing, shown in Fig.3. In (a) and (b), only for $S_3$, the high rank exists. But after the first failure, the high rank is totally deleted. And from the third trial and fourth trial, the low rank for $S_1$ and $S_2$ are emerged and extends with trials respectively. Here we can analyze the situation of $S_1$ as an example. From non-failure to second failure case, the customer can only use PIN to access $S_1$ service with any upper threshold selection ($\Omega > 0.5$). But from third failure to fifth failure, the upper threshold is decreased as 0.65, 0.6, 0.59, 0.51 approximately, which implies that if the custom have five opportunities to stay in medium rank, he can only beg the host setting the upper threshold just a little above the smallest value 0.5. The force-crack is forbidden evidently.

## V. CONCLUSION

This paper proposes a trust model to protect the user's security. The core is to build a service classification estimation table. A fuzzy mathematical method is exploited here. The subjective judgements are quantized into weights. From the model, service classification can mainly evaluate the user's trustiness. In the experiment, thresholds selection is important. Also, the authentication history affects the thresholds of trust region, which makes the trust scheme stricter. The penalty coefficient is shown to be effective for force-crack forbidden.

ACKNOWLEDGMENT

This work was supported by the Fundamental Research Funds for Central Universities: DUT10ZD110.